\documentclass[12pt]{revtex4-2}
\usepackage{amssymb,graphicx,xspace,color}
\setcitestyle{super}

\usepackage{natbib}
\usepackage{lineno}

\begin{document}

\textbf{\LARGE{Fermi surface evolution in Weyl semimetal t-PtBi$_2$ probed by transverse transport properties}}\\
\\
\\F. Caglieris$^{1}$, M. Ceccardi$^{1,2,3}$, D. Efremov$^{3}$, G. Shipunov$^3$, I. Kovalchuk$^{3,7}$, S. Aswartham$^{3}$, A. Veyrat$^{3}$, J. Dufouleur$^{3,5}$, D. Marré$^{1,2}$, B. B\"{u}chner$^{3,4,5}$ \& C. Hess$^{3,5,6}$\\
\\
$^{1}$\textit{CNR-SPIN, 16152 Genova, Italy}\\
$^{2}$\textit{University of Genoa, 16146 Genova, Italy}\\
$^{3}$\textit{Leibniz-Institute for Solid State and Materials Research IFW-Dresden, 01069 Dresden, Germany}\\
$^{4}$\textit{Institut f\"{u}r Festk\"{o}rperphysik, TU Dresden, 01069 Dresden, Germany}\\
$^{5}$\textit{Center for Transport and Devices, TU Dresden, 01069 Dresden, Germany}\\
$^{6}$\textit{Fakult\"{a}t f\"{u}r Mathematik und Naturwissenschaften, Bergische Universit\"{a}t Wuppertal, 42097 Wuppertal, Germany}\\
$^{7}$\textit{Kyiv Academic University, 03142 Kyiv, Ukraine}
\\
\\
\\

\textbf{Abstract}\\

\textbf{The combination of non-trivial topology and superconductivity opens to novel quantum devices. The discovery  of intrinsic materials where such properties appear together represents a frontier in modern condensed matter physics. Trigonal PtBi$_2$ has recently emerged as a possible candidate, being the first example of superconducting type-I Weyl semimetal. However, several aspects of this promising compound still need to be unveiled, concerning its complicated band structure, the actual role of Weyl points in determining its electronic properties and the nature of the superconducting transition. In this work, we experimentally investigated a t-PtBi$_2$ single crystal by means of the Hall and Nernst effects. In particular, we revealed a change of regime in its electronic properties, which is compatible with a temperature and magnetic field evolution of hole-like pockets in the Fermi Surface.} \\

\textbf{Introduction}\\

In the search for advanced materials for novel quantum applications \cite{qi2011topological, sato2017topological, sharma2022comprehensive}, layered trigonal PtBi$_2$ (t-PtBi$_2$) has recently deserved many attentions, thanks to the plethora of fascinating properties it exhibits ranging from non-trivial topology to superconductivity\cite{PhysRevMaterials.4.124202,PhysRevB.97.035133, Feng:2019aa, Gao:2018aa, Nie:2020aa, veyrat2023berezinskii, kuibarov2024evidence, schimmel2023hightc, apxr.202400150}. From the topological point of view, it is a semimetal characterized by broken inversion symmetry, which supports the formation of Weyl points in the bulk \cite{veyrat2023berezinskii, armitage2018weyl}. In addition to that, a remarkable spin-orbit coupling is responsible for a strong Rashba-like spin splitting \cite{Feng:2019aa} and the possible formation of triply degenerated points \cite{Gao:2018aa}. Moreover, in the monolayer limit multiple topological insulator states have been demonstrated \cite{Nie:2020aa}.
However, the most controversial aspect of this material regards its superconducting phenomenology \cite{PhysRevMaterials.4.124202, veyrat2023berezinskii, kuibarov2024evidence, schimmel2023hightc}.  On the one hand, transport measurements on bulk single crystals 
revealed a low superconducting critical temperature ($T_C$) of about 0.6 K \cite{PhysRevMaterials.4.124202}, while Andreev spectroscopy revealed a $T_C$ up to 3.5 K at the interface of point contact\cite{10.1063/10.0014014}. On the other hand, signatures of Berezinskii Kosterlitz-Thouless (BKT) transition have been detected in flakes with thickness up to tens of nanometers \cite{veyrat2023berezinskii}, supporting the two-dimensional (2D) nature of the superconductivity. In addition, angle resolved photoemission spectroscopy (ARPES) revealed that superconductivity appears in the surface states \cite{kuibarov2024evidence}, evidencing the first example of superconducting Fermi arcs. Interestingly, ARPES estimations of the superconducting gap indicates values of 1.4±0.2 meV or 2±0.2 meV (depending on the surface termination), corresponding to critical temperatures of 8±2 K and 14±2 K, respectively. Very recent scanning tunneling spectroscopy (STM) experiments, not only confirmed the presence of surface superconductivity, but provided a strong evidence of superconducting gaps on the surface reaching the surprisingly large value of 20 meV, roughly corresponding to Tc = 130 K \cite{schimmel2023hightc}. The superconductivity is robust against out of plane fields up to about 12 T and also exhibits a significant variability within a surface and even stronger from one prepared surface to another, resembling the phenomenology of underdoped cuprates\cite{schimmel2023hightc}. 
The combination of non-trivial topology, multi-band structure and superconductivity in the same system represents an exceptional coincidence in the light of investigating the formation of Majorana fermions in topological superconductors and t-PtBi$_2$ appears to be a serious candidate to this aim. However, several open issues remain, concerning the actual role of the Weyl nodes, the relationship of surface and bulk transport, the mismatch between transport superconducting critical temperature of 0.6 K and the much larger gap size measured through spectroscopic techniques. All of these aspects claim  for further experimental explorations.\\ 
Here, we investigate a t-PtBi$_2$ macroscopic single crystal by means of transverse electric and thermoelectric transport properties, namely the Hall and the Nernst effect. In particular, the latter has been proven to be a powerful technique to investigate the fermiology of unconventional materials \cite{0034-4885-79-4-046502}. In the vicinity of electronic instabilities, it could exhibit spectacular effects, resulting in an extreme sensitivity to Fermi surface (FS) changes, phase transitions and fluctuations of order parameters \cite{0034-4885-79-4-046502,PhysRevB.80.220514,0953-8984-21-11-113101}. In superconducting compounds, it is considered a primary technique for the detection of several elusive phenomena, including superconducting fluctuations, motion of Abrikosov vortices in the mixed state, pseudogap and time-reversal symmetry breaking\cite{0034-4885-79-4-046502, PhysRevB.80.220514, PhysRevB.102.054503, 0953-8984-21-11-113101, wang2006nernst, grinenko2021state, pallecchi2016thermoelectric}. In addition, in systems characterized by non-trivial topology, a finite Berry curvature at the Fermi level is responsible for the so-called anomalous Nernst effect (ANE)\cite{armitage2018weyl, xiao2006berry}, which recently became a distinctive feature for the identification of Weyl states\cite{PhysRevB.98.201107, PhysRevB.100.085111, yang2020giant, sakai2018giant, ikhlas2017large, ceccardi2023anomalous, rana2018thermopower, liang2017anomalous}. 
Although t-PtBi$_2$ does not exhibit any structural or magnetic transition, our measurements revealed a crossover between different regimes, in which its transverse transport properties undergo substantial changes.
\\

\textbf{Experimental} \\

 The electric and thermoelectric transport measurements have been carried out in He-bath using a home-made probe inserted in a Oxford cryostat endowed of a 15-T-magnet. The Hall effect has been measured in a standard four-probe configuration.
 The Nernst effect has been measured by applying a heating power through a 2.7 k$\Omega$ resistive-heater and the temperature gradient has been evaluated using a Chromel-Au-Chromel thermocouple, calibrated in magnetic field. The angular-dependent Nernst effect has been characterized by adapting an uniaxial piezoelectric rotator by Attocube Systems to the top of our transport probe to create a home-made setup, which allows rotations across an angle -30$^{\circ} < \theta < 120 ^{\circ}$.
Since both the Hall and the Nernst effect are odd with respect to $B$, all the measurements have been performed both in negative and positive fields in order to separate the antisymmetric component from the spurious background.\\
To study the electronic structure, we performed full-relativistic density functional theory (DFT) calculations using the full-potential local-orbital  FPLO package \cite{PhysRevB.59.1743}. For the exchange correlation term we used the generalised gradient approximation (GGA) \cite{PhysRevLett.78.1396}. A k-mesh of 12×12×12 k points in the whole Brillouin zone was employed. The experimental crystal structure of Ref. \cite{PhysRevMaterials.4.124202} was used for the calculations.
\\ 

\textbf{Results} \\

Figure 1a shows the magnetic-field dependence of the Hall resistivity $\rho_{xy}$ at several temperatures. A departure from a standard linear-in-field dependence is observed, with multiple sign changes both as a function of $T$ and $B$, indicating a competition between hole-like and electron-like bands. Figure 1b presents the same data normalized to $B$ as function of the temperature at different constant fields. Interestingly, distinct regimes can be clearly identified. At high temperatures $\rho_{xy}B^{-1}$ is always positive and almost constant or slightly increasing with decreasing $T$. Below a certain temperature $T^*$, which varies in the range 40-100 K depending on the applied magnetic field, $\rho_{xy}B^{-1}$ drops and eventually changes sign. Finally, at low temperatures $\rho_{xy}B^{-1}$ undergoes an upturn, with a turning temperature which increases with $B$ (up to about 20 K for $B$=14 T).\\
Figure 2 shows the evolution of the magnetic-field dependence of the Nernst coefficient $S_{xy}$ with the temperature $T$. As in case of the Hall coefficient, it is possible to distinguish different regimes depending on the temperature range. 
At $T$=100 K, 150 K and 200 K, $S_{xy}$ is positive and linear in high magnetic fields, while it undergoes an anomalous deviation from the linearity at low fields, when approaching $B$=0 T (Figure 2c). In the mid-$T$ region (15 K$<T<$ 50 K) an additional negative contribution appears (confirmed also in a second sample presented in the SI), causing a local minimum located at low fields (Figure 2b). Interestingly, at $T$ = 30 K, 20 K and 15 K, such additional contribution induces a sign change, while at $T$= 50 K the curve turns again completely positive, with the negative term resulting in a broadened feature. At high magnetic fields, the curves at $T$= 15 K and 20 K exhibit a departure from the standard linear $B$-dependence, with a tendency to saturation, which becomes more evident in the low-$T$ curves presented in Figure 2a. At $T$=10 K and 5 K, the negative contribution observed at intermediate temperatures disappears and $S_{xy}$ turns again completely positive.\\
If one expresses $S_{xy}$ in terms of transport coefficients, it is easy to obtain the general expression $S_{xy}=\frac{\alpha_{xy}}{\sigma_{xx}}-S_{xx} \tan \theta_H$ \cite{0953-8984-21-11-113101}, where $\alpha_{xy}$ is the off-diagonal term of the Peltier tensor, $S_{xx}$ is the standard thermopower, and $\tan \theta_H=\frac{\sigma_{xy}}{\sigma_{xx}}$ is the tangent of the Hall angle, which directly links $S_{xy}$ to the Hall conductivity. The formula, which is sometimes called Sondheimer cancellation, is expected to yield zero in one-band systems, where the Hall angle is independent of energy. A finite $S_{xy}$ can thus be considered a measure of the balance between off-diagonal Hall and Peltier coefficients. Interestingly, such expression of $S_{xy}$ holds in case of multi-band transport \cite{0953-8984-21-11-113101}, and the term $-(S_{xx}/T)\tan \theta_H$ can be experimentally derived.
Figure 3 shows the comparison between $S_{xy}/T$ and $-(S_{xx}/T)\tan \theta_H$ at several temperatures. At $T$=100 K $S_{xy}$ is dominated by $\alpha_{xy}/(T\sigma_{xx})$, while $-(S_{xx}/T)\tan \theta_H$ is negligible. In the mid-$T$ regime, namely $T$= 20, 30 and 50 K, $-(S_{xx}/T)\tan \theta_H$ plays instead a crucial role in determining the Nernst signal and it is evidently responsible for the unusual negative contribution in $S_{xy}/T$.
Interestingly, at 20 K, $-(S_{xx}/T)\tan \theta_H$ almost overlaps to $S_{xy}/T$ (Figure 3 b) and $\alpha_{xy}/(T\sigma_{xx})$ has a minimum. At $T$=10 K, both $S_{xy}/T$ and $-(S_{xx}/T)\tan \theta_H$ turn completely positive and $\alpha_{xy}/(T\sigma_{xx})$ strongly increases in amplitude (Figure 3 a).\\
The crossover of the Nernst coefficient from a mid-$T$ to a low-$T$ regime becomes even more evident by investigating what happens to $S_{xy}$ when $B$ is tilted of a $\theta$-angle with respect to the sample $c$-axis, according to the sketch in the Inset of Figure 4 a.  Figure 4 a, presents the angular variation of $S_{xy}$ with $|B|$=15 T at three different temperatures $T$= 10 K, 20 K and 50 K. In the framework of the linear response theory, the Nernst effect is proportional to the component of the magnetic field parallel to the $c$-axis, namely $B_\perp$ = $B$ cos($\theta$). Our data are closer to the expected cos($\theta$)-like only at $T$=50 K, while at $T$=20 K and 10 K, the $\theta$-dependence assumes an unconventional trend. In particular, at $T$=10 K $S_{xy}(\theta)$ becomes non-monotonic and develops two maxima around $\theta = 0^\circ$ and $\theta = 25^\circ$ (mirrored at $\theta = -25^\circ$ ). At $T$=20 K the peak at $\theta = 0^\circ$ appears broadened and that around $\theta = 25^\circ$ becomes barely visible. Figure 4 b-d summarizes in color-plots the behavior of $|S_{xy}|$ as a function of $|B|$cos($\theta$) and $\theta$. The anomalous features observable at $T$= 10 and 20 K appear to be triggered by the temperature and enhanced by the applied field.\\

\textbf{Discussion} \\

The microscopic origin of the observed crossover between different regimes in transverse transport properties of t-PtBi$_2$ remains elusive, since it is not accompanied by evident structural or magnetic phase transitions. This leaves room to a pure electronic mechanism. Generally, strong changes in the Hall resistivity indicate substantial changes in the Fermi surface.
For instance, a drop in the Hall resistivity accompanied by a sign change has been observed in systems characterized by a restructuring in the electronic bands close to the Fermi level, caused by charge density waves or a pseudo-gap-like effect, namely a depletion of electronic states at the Fermi level below a certain temperature. In $2H$-TaSe$_2$, for example, a drop in the Hall resistivity occurs in concomitance of the development of a charge density wave\cite{PhysRevLett.100.236402}, while in several high temperature superconductors (HTCS) a drop and a sign change in $\rho_{xy}$ accompanies the formation of a pseudo-gap above $T_c$ \cite{LeBoeuf2007}. In addition, YBCO also exhibits a sign change of the Nernst coefficient at the onset of the pseudogap phase\cite{PhysRevB.97.064502}, with a phenomenology which partially resembles our observations in t-PtBi$_2$.
The analogy with the pseudo-gap formation in HTCS is interesting in the light of the recent reports about signatures of surface superconductivity by ARPES and STM spectroscopy in PtBi$_2$\cite{ kuibarov2024evidence,schimmel2023hightc}, with indication of reduced density of states in tunneling experiments, already starting at relatively high temperatures\cite{schimmel2023hightc}. However, while intriguing, the hypothesis of a pseudo-gap in t-PtBi$_2$ requires further investigation at this stage, given the absence of clear signatures of high-temperature superconductivity in transport properties.\\ 
FS reconstructions and concomitant changes in the Hall resistivity may also occur without involving any superstructure \cite{wu2015temperature, PhysRevB.93.024503}. This has been observed for instance in LiFeAs, where ARPES revealed a change of the electronic structure at some pockets of the Fermi surface up to 40 meV, in absence of structural or density wave transitions \cite{PhysRevB.93.024503}. To develop this hypothesis, we can consider that, as a general statement, in multi-band systems, a sign change in the Hall resistivity is assigned to a change of the dominant charge carrier type. 
Intuitively, the observed drop in the Hall resistivity from positive to negative values could be explained through an increase of electron-like carriers or a depletion of hole-like carriers at the Fermi level. 
In our case a substantial variation of hole-type carriers is more probable because in t-PtBi$_2$ holes generates smaller Fermi pockets with respect to electrons \cite{Gao:2018aa} and tiny Fermi pockets are much more susceptible to temperature (and magnetic field) changes of the chemical potential with respect to larger ones.
This scenario is corroborated by our Fermi Surface calculations. In Figure 5 a we present the Fermi surface of t-PtBi$_2$ calculated in the frame of the DFT theory, which is consistent with previous reports\cite{Gao:2018aa, veyrat2023berezinskii}. Figures 5 b-f show the change of the FS for tiny shifts of the Fermi Energy by 2 meV steps, from -14 meV to -22 meV. Such evolution could simulate, for instance, a temperature-dependent chemical potential. Interestingly, a minimal change of few meV is already capable to induce substantial changes in the FS, including a clear Lifshitz transition. In particular, the \textit{finger-like} structures, which link the top and the bottom part of the FS, are connected in Figure 5 e and 5 f (corresponding to -20 meV and -22 meV) and get disconnected in Figure 5 d (corresponding to -18 meV). Remarkably, such \textit{finger-like} structures are formed by hole-like bands. If we assume that the actual Fermi level of our t-PtBi$_2$ is located in proximity of the Lifshitz transition, a shift of the chemical potential of few meV can substantially change the hole-type carriers altering the balance between hole- and electron- transport, sensitively probed by $\rho_{xy}$. In addition, we tested the effect of a magnetic field of the FS imposing a finite magnetization of 0.01 $\mu_B$ in our calculations, which roughly corresponds to the effect of a 20 T external magnetic field coupled to the spin degree of freedom. Interestingly, there is a slight variation to the FS in correspondence of the \textit{finger-like} structures, indicating that the presence of $B$ could induce a minimal energy shift in the Lifshitz transition (See SI for further information).\\
Let's now consider the Nernst phenomenology. 
In the previous section we decomposed the Nernst coefficient into two parts. As we already mentioned, the $-(S_{xx}/T)\tan \theta_H$ component is strictly related to the behavior of the Hall resistivity and determines the unusual negative contribution observed in $S_{xy}$ in the mid-$T$ region. 
The term $\alpha_{xy}/(T\sigma_{xx})$ is more subtle because $\alpha_{xy}/T$ is proportional to the derivative of the Hall conductivity with respect to the chemical potential $E_F$, namely $\alpha_{xy}/T \propto -(d\sigma_{xy}/ dE_F)$. 
This means that a finite $\alpha_{xy}$ typically requires tiny Fermi pockets, highly sensitive to changes of $E_F$, while large portions of FS do not give substantial contributions\cite{0953-8984-21-11-113101}. 
Our FS calculations suggest that in t-PtBi$_2$ hole pockets are much more susceptible to variation of the chemical potential with respect to electron ones. Hence, we can assume that $\alpha_{xy}$ is dominated by hole-like carriers, reading $\alpha_{xy}/T \propto -d\sigma_{xy}^h/dE_F$, where $\sigma_{xy}^h=en_h\mu^2_hB$, $n_h$ and $\mu_h$ are the Hall conductivity, the density and the mobility of the hole-like carriers, respectively. The derivative $d\sigma_{xy}^h/dE_F$ can be decomposed into two parts, obtaining $d\sigma_{xy}^h/dE_F=eB(\mu_h^2\frac{dn_h}{dE_F}+2n_h\mu_h \frac{d\mu_h}{dE_F}$). If the FS is highly susceptible to variation of the chemical potential, it is reasonable to suppose that the term proportional to $dn_h/dE_F$ is the dominant one, simplifying the expression in $\alpha_{xy}/T \propto -eB\mu_h^2dn_h/dE$.

In case of free carriers $E_F(T)-E_F(T=0)=\Delta E_F(T) \propto-T^2$, meaning that $\alpha_{xy} \propto B\mu_h^2 dn_h/dT$. In Figure 3 f we report the behavior of $\alpha_{xy}/\sigma_{xx}$ normalized to the magnetic field $B$: we can distinguish two evident regimes, with a crossover at around 20 K. The positive $\alpha_{xy}B^{-1}/\sigma_{xx}$ for $T>$20 K is consistent $dn_h/dT>$0, possibly corresponding to a hole carrier depletion below a certain temperature. 
The explosion of $\alpha_{xy}B^{-1}/\sigma_{xx}$ at 10 K could be instead related to a strong increase of the hole mobility at low temperatures (amplified by the $\mu^2$-dependence of $\alpha _{xy}$), which also explains the low-temperature upturn of $\rho_{xy}B^{-1}$ in Figure 1 b. Interestingly, an abrupt enhancement of the carrier mobility in the same temperature region has been also reported in other semimetals, such as WTe$_2$ \cite{Pan2022}.\\
The temperature evolution of the electronic properties of t-PtBi$_2$ becomes even more evident from the colored plots in Figure 4 b-d, which show a dramatic change of the angular-dependence of Nernst coefficient from 50 K to 10 K. In particular, in Figure 4 b, it is interesting to notice the presence of Shubnikov de Haas quantum oscillations in $|S_{xy}|$, appearing at $-30^\circ<\theta<30^\circ$ and high magnetic fields. We observed that the oscillation frequency does not substantially change with the angle, while their amplitude gets strongly modulated, with maxima located around $\theta\approx\pm 20^\circ$ (See SI). On the one hand the constant frequency suggests that the oscillations are generated by relative isotropic pockets, while, on the other hand, the modulation of the amplitude points towards a $k$-space dependence of the mobility of the charge carriers. In fact, the within the Lifshiz-Kosevic model, the oscillation amplitude is damped by the Dingle factor $R_D=e^{-\frac{\pi m^*}{e B \tau}}$, which depends on the effective mass $m^*$ and the qunatum scattering time $\tau$ (see Supplementary Information).\\
The tensor nature of the carrier mobility at low temperatures can also explain the broadened maxima in the $\theta$-dependence of Figure 4 a, since the magnitude $S_{xy}$ directly depends on the mobility itself. This argument has been already proposed to explain the angular dependence of the magnetoresistance in pure Bi single crystals \cite{collaudin2015angle}.\\
Hence, a possible scenario that emerges from the analysis of our transverse transport properties includes two main ingredients. First of all, the sensitivity of hole-like pockets to minimal variation of the chemical potential, which make them strongly susceptible to temperature and magnetic field changes. This can justify the drop and the sign change in the Hall resistivity below $T^*$ as well as the temperature variation of $\alpha_{xy}$. 
Secondly, a $k$-space dependence of the carrier mobility and its considerable increase at low temperatures, which can explain the unusual $\theta$-dependence of $S_{xy}$ and its rapid enhancement below 20 K.\\ 
Finally, we would like to comment on the possible signature of non trivial topology in our transport measurements. In facts, since Weyl nodes are a natural source of Berry curvature, the appearance of non-linear-in-field Nernst effect in topological semimetals has been often attributed to an ANE contribution, even in non-magnetic compounds such as Cd$_3$As$_2$ \cite{liang2017anomalous} or TaP \cite{PhysRevB.98.201107}. Although a similar mechanism cannot be completely excluded a priori in PtBi$_2$, our material presents substantial differences with respect to the mentioned cases. If we consider the curves at 5 K and 10 K presented Figure 1 a, $S_{xy}/T$ shows a tendency to a slow saturation with $B$ rather then the step-like behavior followed by a plateau, which is considered the fingerprint of the ANE\cite{liang2017anomalous}. Such smooth $B$-dependence in t-PtBi$_2$ can be easily attributed to the crossover from a low-field limit to a high field regime within a multi-band picture \cite{PhysRevB.98.155116}.  
Secondly, Weyl nodes in t-PtBi$_2$ are predicted to be located at about 47 meV above the Fermi level \cite{kuibarov2024evidence}, which may be too far to make them dominate the Nernst effect with an anomalous component. Indeed, the ANE is weighted by the entropy density which is sizable only in a $k_BT$ range around the Fermi level \cite{PhysRevB.100.085111}. In addition, Tewari et al. evaluated the Nernst effect within a lattice model for inversion-symmetry breaking Weyl semimetals, observing that in general the normal component should be dominant on the anomalous one \cite{PhysRevB.96.195119}. 
The only feature in our measurements that resembles an anomalous component is the low-field anomaly observed in the $S_{xy}/T$ vs $B$ curves at $T$= 100 K, 150 K and 200 K (Figure 2 c). This anomaly is puzzling since it disappears at lower temperatures and it has no counterpart in other transport properties, such as the Hall resistivity. A further analysis of this contribution is proposed in the SI, but a deeper experimental and theoretical investigation will be matter of future works.\\

In conclusion, we measured the Hall and the Nernst effects in a t-PtBi$_2$, a system which attracted a great attention due to the combination of non-trivial topology and surface superconductvity. Our transport study revealed a temperature cross-over between different regimes, which is compatible with a non-trivial evolution of t-PtBi$_2$ electronic properties, despite the absence of structural or magnetic transitions. Beside more exotic explanations, including the formation of pseudo-gap like features, possibly in connection with the observed surface superconductivity, we believe that the most plausible picture for the observed phenomenology consists in a peculiar temperature and magnetic field dependence of its multifaceted Fermi Surface. We checked the consistency of such hypothesis through ab initio calculations, which revealed the presence of hole-like pockets, that are strongly susceptible to minimal shifts of the chemical potential.\\
Although our study does not reveal any evident contribution related to non trivial topology or superconductivity, our work draws attention on the high tunability of some portions of t-PtBi$_2$ Fermi Surface, feeding the idea that different external parameters, such as magnetic fields, temperature or strain could easily trigger changes in its electronic properties.

\textbf{Acknowledgments}
This work has been supported by the Deutsche Forschungsgemeinschaft (DFG)
through SFB 1143, Project-ID No.247310070 (CH). DE acknowledges support of DFG through the project N 529677299 and 455319354. SA acknowledges financial support by the Deutsche Forschungsgemeinschaft (DFG) through the Grant AS 523/4-1. This work was supported by the Deutsche Forschungsgemeinschaft (DFG, German Research Foundation) under Germany’s Excellence Strategy through the Würzburg-Dresden Cluster of Excellence on Complexity and Topology in Quantum Matter—ct.qmat (EXC 2147, project ids 0242021 and 392019),and by the Leibniz Association through the Leibniz Competition (JD). \\

\textbf{Author contributions} FC, MC and AV performed all the transport measurements. DE performed the Fermi Surface calculations. GS and SA synthesized the samples. FC, DE, JD, CH and BB interpreted the experimental results. FC, DM, CH and BB supervised the project.\\

\textbf{Data Availability} The data that support the findings of this study are available from the corresponding author upon reasonable request. \\

\textbf{Competing Interests} The authors declare that they have no competing financial interests.\\

\textbf{Correspondence} Correspondence should be addressed to F. Caglieris~(email: federico.caglieris@spin.cnr.it).\\

\newpage

\newpage

\begin{figure}[!t]
\includegraphics[width=\columnwidth]{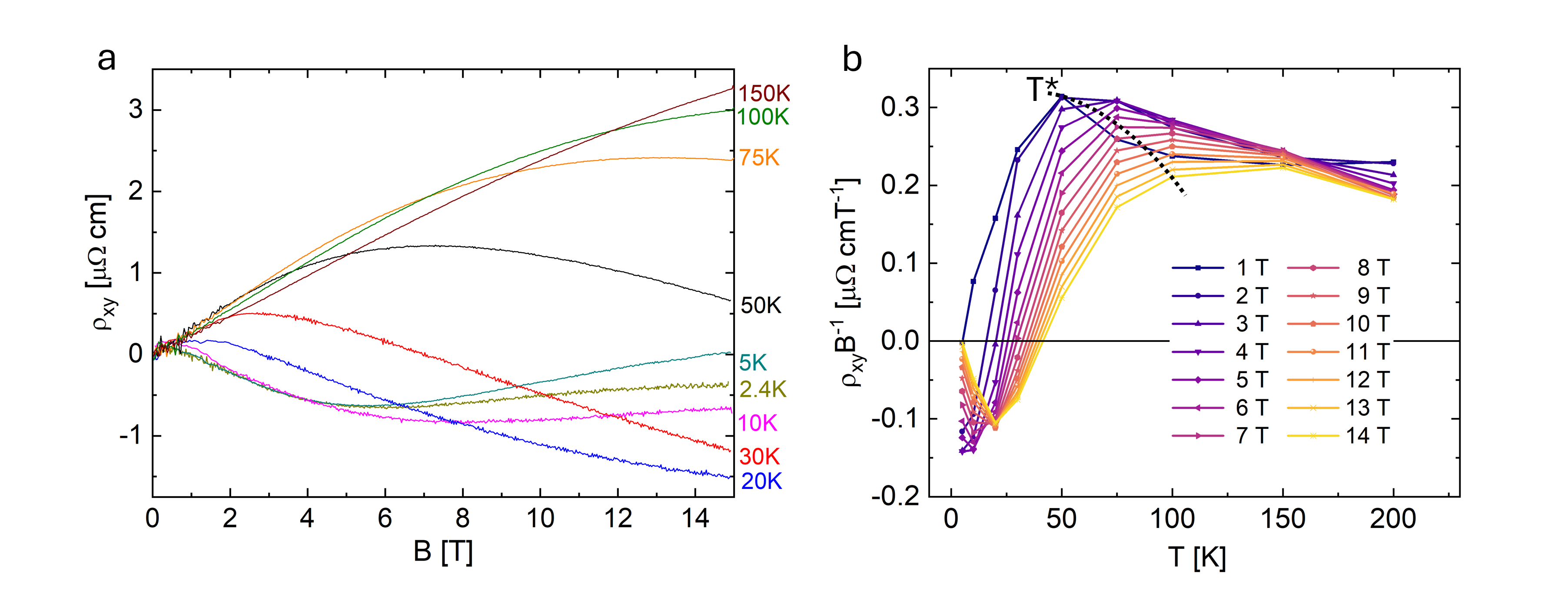}
\caption{\label{Figure} a) $B$-dependence of the Hall resistivity $\rho_{xy}$ of the PtBi$_2$ single crystal at several temperatures. b) $T$-dependence of the Hall resistivity $\rho_{xy}$ of the PtBi$_2$ single crystal at several magnetic fields}
\end{figure}

\begin{figure}[!t]
\includegraphics[width=\columnwidth]{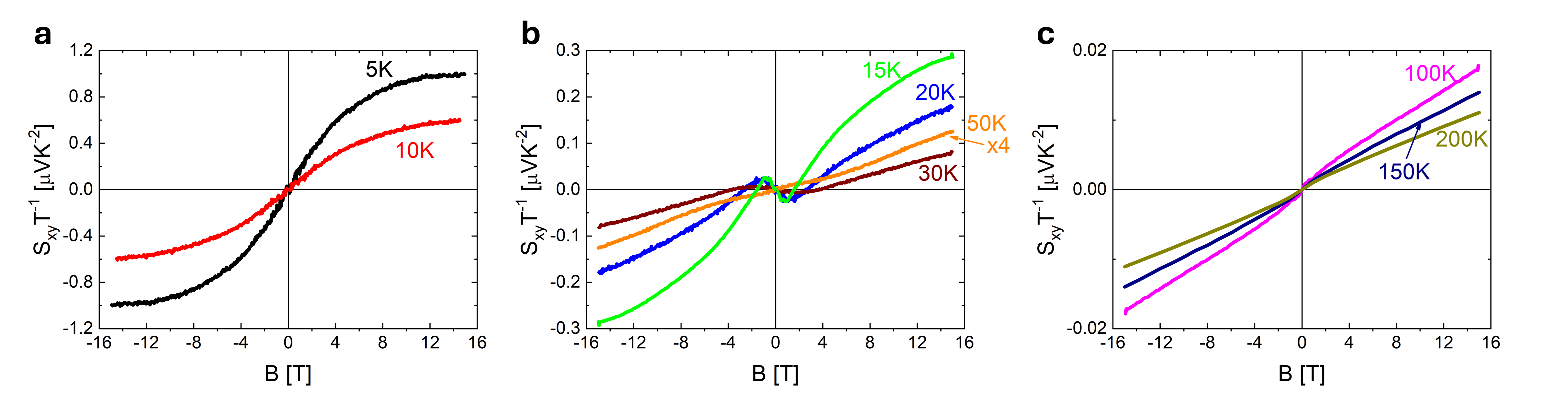}
\caption{\label{Figure} $B$-dependence of $S_{xy}T^{-1}$ in the a) low-temperature, b) mid-temperature, c) high-temperature regimes.}
\end{figure}

\begin{figure}[!t]
\includegraphics[width=\columnwidth]{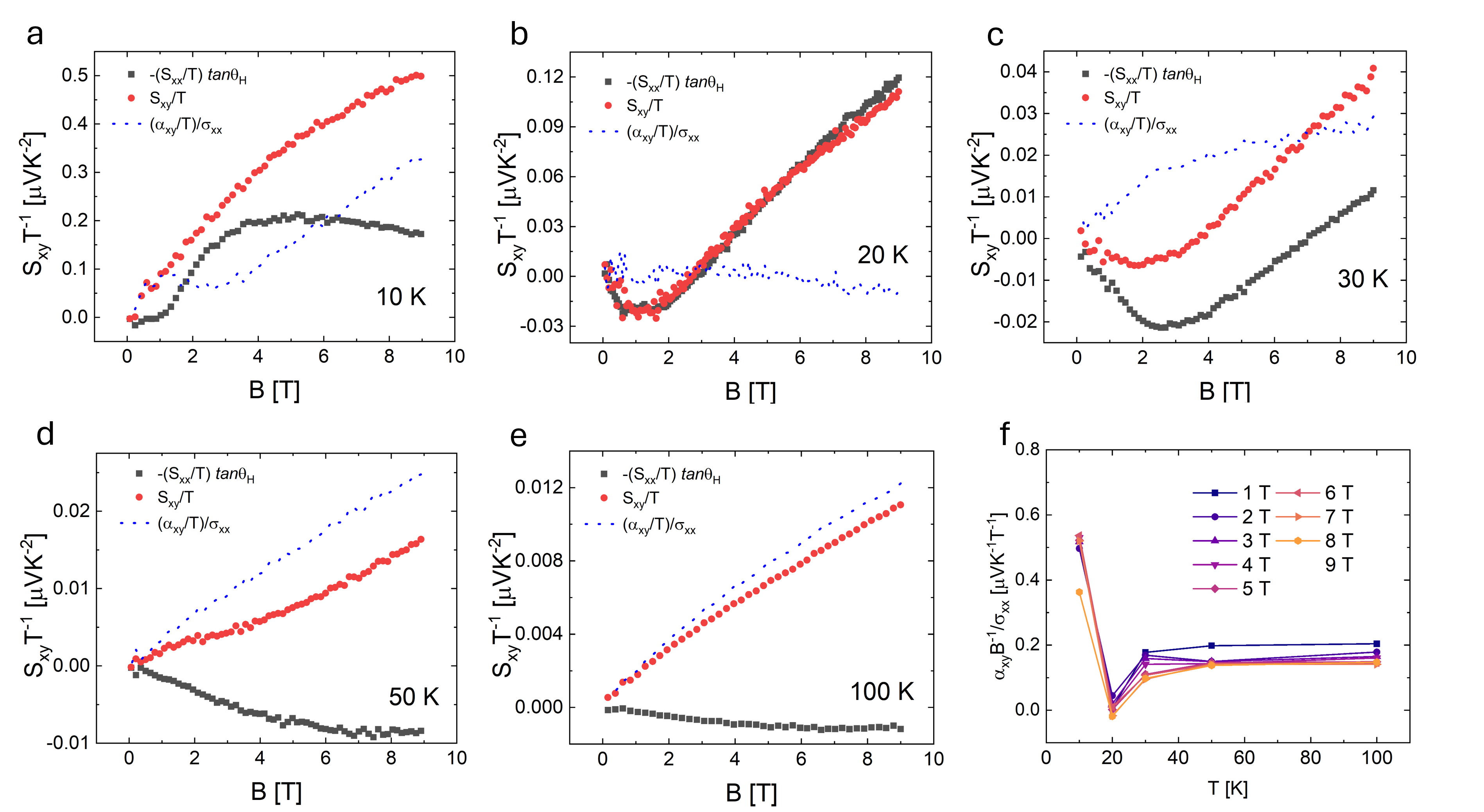}
\caption{\label{Figure} $B$-dependence of $S_{xy}T^{-1}$ (red symbols), $-(S_{xx}/T)\tan \theta_H$ (black symbols) and $\frac{(\alpha_{xy}/T)}{\sigma{xx}}$ (blue dots) at a) T= 10 K, b) T= 20 K, c) T= 30 K, d) T= 50 K and e) T= 50 K. f) $T$-dependence of the $\alpha_{xy} B^{-1}/\sigma_{xx}$ for different magnetic fields from 1 T up to 9 T.}
\end{figure}

\begin{figure}[!t]
\includegraphics[width=\columnwidth]{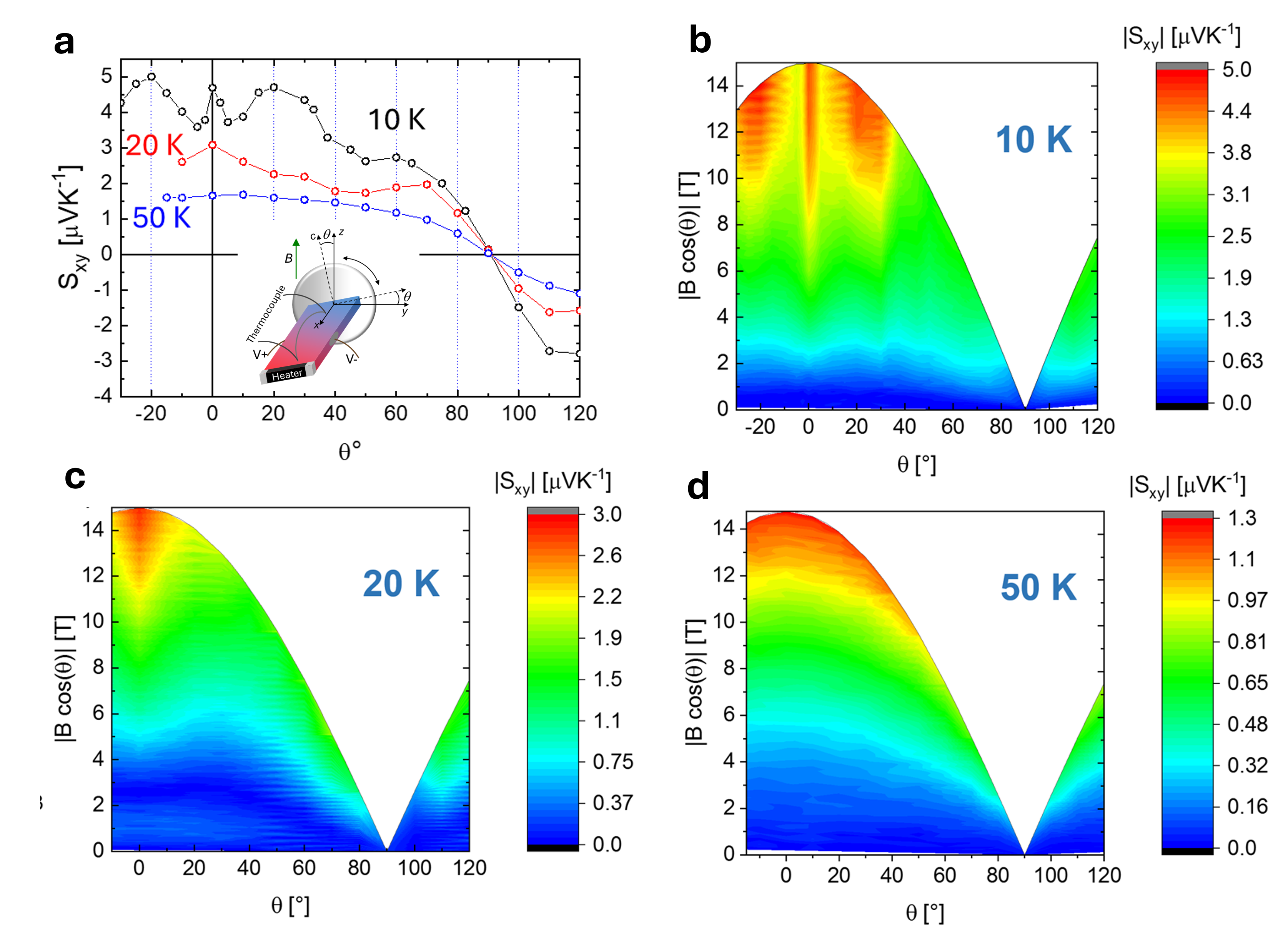}
\caption{\label{Figure}a) $\theta$-dependence of $S_xy$ at $B$= 15 T and $T$= 10 K, 20 K and 50 K. Inset. Schematic of the measurement setup. A heat-gradient is applied along the $x$-axis and measured through a Chromel-Au-Chromel thermocouple. The magnetic field $B$ is applied along the $z$-axis and the Nernst voltage is measured across two electrodes positioned along the $y$-axis. The sample is free to rotate around the $x$-axis. b) - d) Module of the Nernst coefficient ($|S_{xy}|$) as a function of the angle ($\theta$) and the the projection of magnetic field along the z-direction ($|B \cos \theta|$) at b) $T$=10 K, c) $T$=20 K and d) $T$=50 K. The color scale represents the amplitude of $|S_{xy}|$}
\end{figure}

\begin{figure}[!t]
\includegraphics[width=\columnwidth]{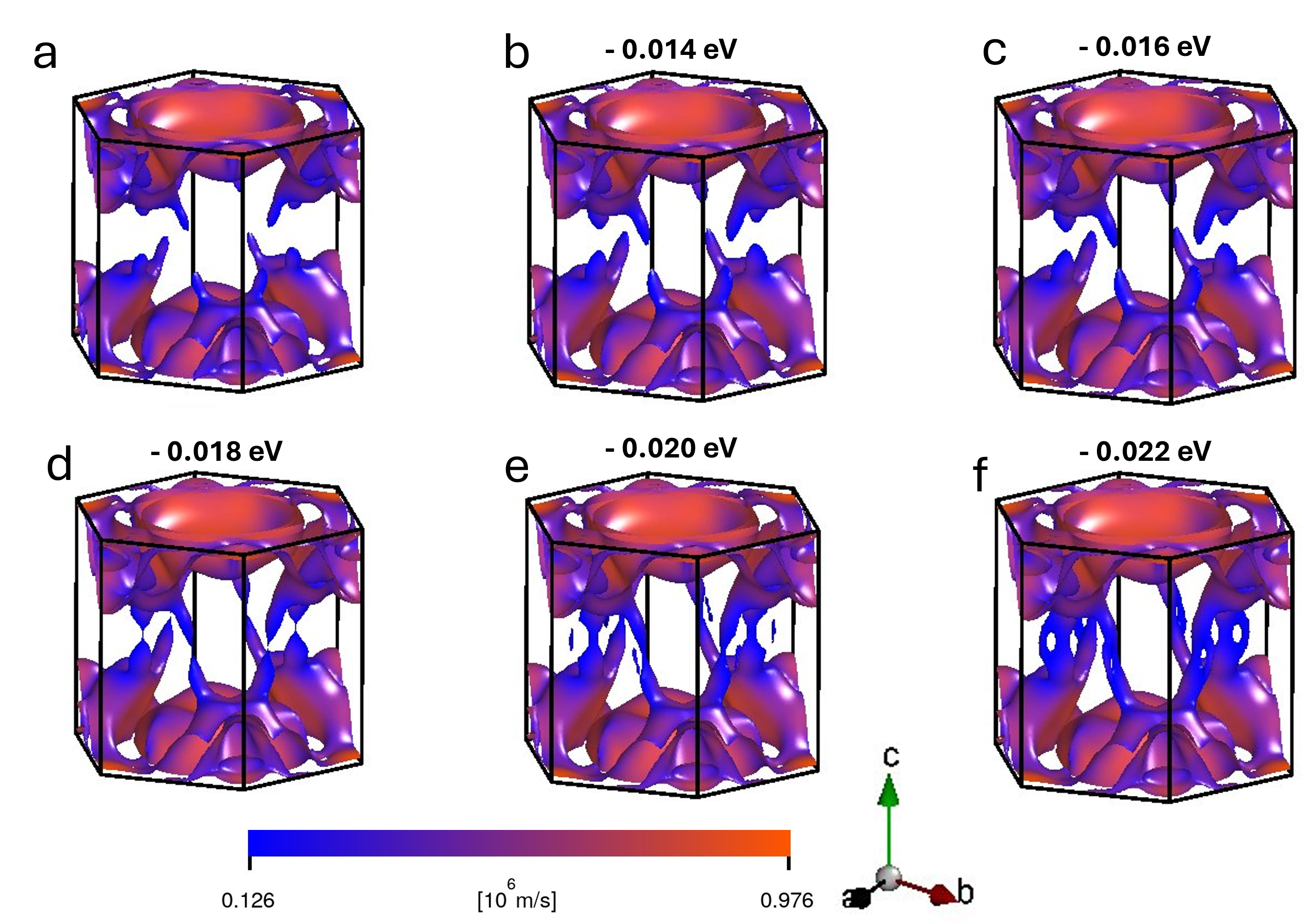}
\caption{\label{Figure} Evolution of t-PtBi$_2$ Fermi surface as a function of different values of the Fermi energy, ranging between -0.014 eV and -0.022 eV (panels b - f) evaluated with respect to the reference in panel a. A Lifshitz transition is evident between -0.020 and -0.018 eV.}
\end{figure}

\end{document}